# ASSORTATIVE MIXING IN CLOSE-PACKED SPATIAL NETWORKS

**Deniz Turgut, Ali Rana Atilgan, Canan Atilgan**

Faculty of Engineering and Natural Sciences, Sabanci University, 34956 Istanbul, Turkey

**Corresponding author:** Canan ATILGAN; *e-mail*: canan@sabanciuniv.edu; *telephone*: +90 (216) 4839523; *telefax*: +90 (216) 4839550

*All computer programs used in the analyses are available upon request.

**Abstract**

A general relation for the dependence of nearest neighbor degree correlations on degree is derived. Dependence of local clustering on degree is shown to be the sole determining factor of assortative versus disassortative mixing in networks. The characteristics of networks derived from spatial atomic/molecular systems exemplified by self-organized residue networks and block copolymers, atomic clusters and well-compressed polymeric melts are studied. Distributions of statistical properties of the networks are presented. For these densely-packed systems, assortative mixing in the network construction is found to apply, and conditions are derived for a simple linear dependence. Together, these measures (i) reveal patterns that are common to close-packed clusters of atoms/molecules, (ii) identify the type of surface effects prominent in different systems, and (iii) associate fingerprints that may be used to classify networks with varying types of correlations.

# I. Introduction

The study of real life networks, such as the world-wide web [1], internet [2], power-grids [3] and math co-authorship [4], has put forth properties that distinguish them from classical Erdös-Renyi random networks [5]. The variety of degree distributions and other statistical measures that emerge has heightened the interest in complex networks. With the proposition of algorithms by Watts-Strogatz [3] and Barabási-Albert [6] to generate real life-like networks, this area has been investigated extensively [7-8]. The classification of networks is mostly based on measures such as degree distributions, average clustering, and average path length [9-10]. Recently, spectral properties of networks gained attention since the distribution of eigenvalues characterize several aspects of the network such as algebraic connectivity and bipartiteness [11-13]. Although there may be different graphs structures with identical Laplacian spectra that defines the network, they often show similar characteristics in terms of network parameters [14]. Several heuristic algorithms are proposed to generate networks from their spectra [15].

In recent years, proteins were investigated as networks, by taking the amino-acids as nodes. Termed as residue networks (RN), edges between neighboring nodes are represented by their bonded and non-bonded interactions [16-19]. Several studies have shown that residue networks have small-world topology [16, 20-22], characterized by their logarithmically scaling average path lengths with network size, despite displaying high clustering. Further studies also utilized network models for protein structures to predict hot spots [23-26], conserved sites [23-29], domain motions [23-26, 30-31], functional residues [32-35] and protein-protein interactions [36]. The small-world topology of residue networks is established, and various network properties such as the clustering coefficient, path length, and degree distribution are used to account for, e.g. the different fold-types in proteins [27], interfacial recognition sites of RNA [28], and bridging interactions along the interface of interacting proteins [17]. In light of these studies, we expect other self-organized molecular systems of synthetic origin to display similar topology.

In fact, a hierarchical arrangement of the nodes is expected to occur in self-organization of atoms and molecules under the influence of free energetic driving forces. In graph theory, hierarchies have been quantified by the presence of (dis)assortative mixing of their degrees, defined as nodes with high degrees having a tendency to interact with other nodes with (low)high degrees [37]. Analytical and computational models for generating assortatively mixed networks were proposed [38-39]. Newman has shown that assortatively mixed networks percolate more easily and they are more robust towards vertex removal [38, 40]; most social networks are examples of these. We find in this work RN in proteins to also have assortative mixing, although many biological networks such as protein-protein interactions and food webs were found to display disassortative behavior.

It is expected that in networks displaying any degree of correlations, local properties of the constructed graphs will have an effect on the global features. However, a connection between the local and global network properties and the underlying structure of molecular systems has yet to be established. In this study, following a general definition of network descriptors (section II), we derive a relationship relating the nearest neighbor degree correlation of nodes, their degree, and clustering coefficient (section III). After a brief description of the model systems studied (section IV), we show in section V that a linear relationship is valid for two types of self-organized molecular systems: (i) Folded proteins and (ii) block co-oligomers in a solvent that encourages micelle formation. Furthermore, simulated configurations of Lennard-Jones clusters also approximate the findings as well as a simple polymeric system forced into a close-packed structure under extremely high pressure. We next show that model hexagonal close packed structures may be used to reproduce many of the graph properties of the above-mentioned systems. This study is a first step towards using both statistical and spectral characterization in determining the design principles underlying organization of complex molecular networks.

## II. Network Descriptors

An un-weighted simple network can be identified fully via the adjacency matrix (**A**), constructed as

$$A_{ij} = \begin{cases} 1 & \text{if nodes } i \text{ and } j \text{ are connected} \\ 0 & \text{if nodes } i \text{ and } j \text{ are not connected} \end{cases} \qquad (1)$$

Several parameters are defined to classify networks; each can be computed from the adjacency matrix and are considered as either a local or a global parameter. The simplest parameter is the connectivity, $k_i$, of node $i$, also known as the degree;

$$k_i = \sum_{j=1}^{N} A_{ij} \qquad (2)$$

Poisson, Gaussian or Power law degree distributions are observed in many real life networks.

Higher order degree correlations are also of importance and may be utilized to identify more distinguishing features of the network. For instance, second degree correlation of a node $i$, denoted by $k_{nn,i}$, is the average connections of its neighbors and may be written in terms of the adjacency matrix.

$$k_{nn,i} = \sum_{j=1}^{N} \sum_{m=1}^{N} A_{ij} A_{jm} = \sum_{j=1}^{N} A_{ij} k_j \qquad (3)$$

$k_{nn,i}$ is also referred to as nearest-neighbor degree correlation. Normalized third degree correlations ($C_i$), known also as the clustering coefficient, is widely used to characterize the distinctness of networks [3, 6]. It is defined as the ratio of the number of interconnections between a node's neighbors to the number of all its possible connections, i.e.;

$$C_i = \frac{\frac{1}{2}\sum_{j=1}^{N} \sum_{m=1}^{N} A_{ij} A_{jm} A_{mi}}{\frac{k_i(k_i - 1)}{2}} \qquad (4)$$

While $k_i$, $k_{nn,i}$, and $C_i$ are descriptors of local structure, another common parameter used to classify the global structure of networks is the average shortest path length, $L_i$ of a node. Given that the shortest number of steps to reach node $i$ from node $j$ along the network is $L_{ij}$, it is the average number of steps that are traversed from all other nodes to node $i$.

### III. Relationship between $k_{nn}$ and $k$

We expand on the treatment in ref. [41] to derive a general relationship for the nearest-neighbor degree correlation (equation 3) for graphs with non-negligible clustering coefficients. The generating function, $G_0(x)$, for the probability distribution of vertex degrees $k$ is given by,

$$G_0(x) = \sum_{k=0}^{\infty} p_k x^k \qquad (5)$$

where $|x| \leq 1$, $p_k$ is the probability that a randomly chosen vertex on the graph has degree $k$, and its distribution is normalized with $G_0(1) = 1$. The $G_0(x)$ function generates the probability distribution, capturing all the discrete probability values through the derivatives property,

$$p_k = \frac{1}{k!} \frac{d^k G_0}{dx^k}\bigg|_{x=0} \tag{6}$$

The $n^{th}$ moment of the distribution can thus be calculated from

$$\langle k^n \rangle = \sum_k k^n p_k = \left[\left(x\frac{d}{dx}\right)^n G_0(x)\right]_{x=1} \tag{7}$$

In particular, the average degree of a vertex is $\langle k \rangle = z = \sum_k k p_k = G'_0(1)$.

If one randomly chooses $m$ vertices from a graph, than the powers property of the generating function provides a route to generating the distribution of the sum of the degrees of those vertices by $[G_0(x)]^m$.

We define *outgoing edges* from the first neighbors of a randomly chosen vertex as those that connect to vertices that are different from the first neighbors of the originally chosen vertex. It is first necessary to define the generating function for the distribution of the degree of the vertices one arrives at, along a randomly chosen edge. That vertex will be reached with probability proportional to its degree, $kp_k$, so that the normalized distribution is generated by

$$\frac{\sum_k k p_k x^k}{\sum_k k p_k} = x\frac{G'_0(x)}{G'_0(1)} \tag{8}$$

Starting from a randomly chosen vertex and following each of its edges to arrive at the $k$ nearest neighbors, each of the vertices arrived at will have outgoing edges that is given by the degree of that vertex less the edge that one arrives along and the backlinks, $b$. The latter are defined as the edges that interconnect the nearest neighbors of the original vertex. Thus, the generating function for the outgoing edges from *each* vertex is,

$$G_1(x) = \frac{\sum_k k p_k x^{k-1-b}}{\sum_k k p_k} \tag{9}$$

Note that $b$ itself depends on $k$.

The number of backlinks, $b$, is given in terms of the clustering coefficient, $C$, around a given node with degree $k$. Using the definition of $C$, with the number of interconnections, $I$, between its first neighbors, $C = I/[k(k-1)/2]$, the average number of backlinks for each of the $k$ neighboring nodes is, $b = 2I/k = C(k-1)$. This will lead to the generating function for outgoing edges as:

$$G_1(x) = \frac{\sum_k k p_k x^{(k-1)(1-C)}}{z} \tag{10}$$

The generating function for the distribution of all outgoing links from the $k$ neighbors of the original node is then obtained from the powers property:

$$G_k(x) = G_1(x)^k = \left[\frac{\sum_k k p_k x^{(k-1)(1-C)}}{z}\right]^k \tag{11}$$

The average number of outgoing links is computed (8) from the first moment of the generating function evaluated at $x = 1$. In general, this leads to

$$G'_k(1) = k\left(\frac{\langle k^2 \rangle}{z} - 1\right) - \frac{k}{z}\left[\sum_k C k p_k (k-1)\right] \tag{12}$$

$k_{nn}$ is the nearest neighbor correlations, defined as the total number of neighbors of a given node which emanates from a selected node of $k$ neighbors. Thus, it is the sum of the number of outgoing links per neighbor, the backlinks per neighbor and the link that connects the original node to the first neighbor:

$$k_{nn} = \frac{G'_k(1)}{k} + b + 1 = \left\{\frac{\langle k^2 \rangle}{z} - \frac{1}{z}\left[\sum_k C k p_k (k-1)\right]\right\} + C(k-1) \tag{13}$$

The first term in curly brackets is constant, carrying information on the moments of the distributions, depending on how $C$ is related to $k$. The second term determines the assortative versus disassortative behavior of the network. For example, if $C$ decreases with $k$ as a single exponential, $C \propto \exp(-ak)$, we may get assortative or disassortative mixing depending on the strength of the decay. For the cases of $C \to 0$, one gets uncorrelated networks. On the other hand, for the particular case of a system where $C$ is finite, yet independent of $k$, equation 14 reduces to the simple linear expression:

$$k_{nn} = Ck + \frac{\langle k^2 \rangle (1-C)}{z} \tag{14}$$

with slope $C$ and the intercept depending on the degree distribution. For example, for a Poisson distributed network, e.g. approximated by residue networks constructed from folded protein structures as was shown in [16-17], $p_k = z^k e^{-z}/k!$, the relation takes the form

$$k_{nn} = Ck + (1+z)(1-C) \qquad (15)$$

In this work, we study concentrated atomic/molecular systems which have a weak dependence of clustering coefficient on degree. We shall see that the linear dependence of equation 14 suffices to describe their nearest neighbor degree correlations.

In passing, we note that an algorithm for generating networks with given clustering dependence on degree has previously been proposed [42]. However, the algorithm fixes the average clustering coefficient and has no control over the distribution of clustering for a given degree, while this distribution is crucial in our derivation. Moreover, they impose the constraint for networks to be assortatively mixed in order to construct the desired network with their proposed algorithm.

## IV. Model Systems

**A. Self-organized molecular structures.** In this subsection we describe how the networks are constructed for the two self-organized molecular structures studied in this work.

*Residue Networks (RN):* These networks are formed from experimentally determined protein structures obtained from the Protein Data Bank (PDB) [43]. For the RN calculations we utilize a set of 595 single-chain proteins with sizes between 54-1021 and having a sequence homology less than %25 [44]. This protein set is identical to the set we used in our previous studies [16-17] and is listed as a supplementary file in [17].

Given a protein, each amino-acid is represented by a node that is centered at the position of $C_\beta$ atoms, or the $C_\alpha$ atom in the case of Glycine. Edges are added between two nodes (i.e. $A_{i,j}$ = 1 in equation 1), if they are closer than a selected cutoff distance, $r_c$. We call these constructions residue networks (RN). We use $r_c = 6.7$ Å as in our previous work, which is the distance where the first coordination shells ends, as computed from the radial distribution function (RDF) shown in figure 1. See references [16-17, 45] for more details on the construction of residue networks and the choice of $r_c$.

*Micellar Networks (MN).* Unlike proteins, there is no experimentally available atomistic structure data for self-organized synthetic molecules. We therefore generate such data using Dissipative Particle Dynamics (DPD) simulations. DPD is a coarse grained simulation methodology. The equilibrium morphology of a group of beads is obtained by integrating out the fast motion of atoms. In addition to the random and dissipative forces, the net forces on the beads are soft and repulsive conservative forces. The simulation is carried out by integrating Newton's law of motion. DPD simulations allow reaching much larger length and time scales for macromolecular systems. Thus, morphologies of self-organized systems of large sizes can be studied. Here, we simulate the micelle formation by ABC type oligomers of styrene-co-perfluoroalkylethylacrylate in tetrahydrofuran (F beads). The co-oligomer consists of ten styrene monomers (A beads), seven perfluoroheptane monomers (C beads) and a linker monomer (B bead). The styrene monomers in the co-oligomer have a tendency to interact with the solvent, whereas the fluorinated parts prefer to segregate, thus resulting in micelle formation. The equilibrium morphology depends on the concentration of oligomer in the solution [46]. Force on bead $i$ is given by $\mathbf{f}_i = \sum_{j \neq i}(\mathbf{F}_{ij}^C + \mathbf{F}_{ij}^D + \mathbf{F}_{ij}^R) + \sum_k \mathbf{F}_{ij}^{conn}$, where the respective forces are due to interaction, dissipative and random forces between beads $i$ and $j$, and chain connectivity between bead $i$, its neighbors $k$ along the chain contour. A general overview of the DPD method and parameterization details for this particular system is given in [47].

We report results from systems where the volume fraction, $\nu$, of the oligomers is 0.3, 0.6 and 0.9, respectively. We find that at these concentrations, the triblock co-oligomers self-organize into spherical, cylindrical and lamellar morphologies respectively, as the concentration is increased. Once the organized structures are obtained, we focus on one substructure from the simulated system; e.g. the set of oligomers that form a complete sphere are taken as the structure whose network will be formed. Thus, the spherical structure is made up of 50 chains, the cylindrical structure has 100 chains, and the lamellar structure has 150 chains. Finally, we concentrate on the fluorinated segments of these segments, which have self-organized due to the driving forces inherent to the system beads. By computing the RDFs around these beads, we find that the first coordination shell ends at 1.1 DPD units (see figure 1). We use this cutoff distance to form the network (equation 1) whose properties are studied. Chain connectivity of a copolymer is preserved regardless of the particle separation; i.e. ($i$, $i$+1) connections are always present. Also shown as an inset to figure 1 are sample configurations of spherical, cylindrical and lamellar formations excerpted from oligomer concentrations of $\nu$ = 0.3, 0.6, and 0.9, respectively.

**B. Other atomic/molecular structures.** We also study other densely packed systems of atomic/molecular origin, to investigate the effects of excluded volume and chain connectivity on the observed statistical and spectral properties. To this end, we focus on the structure of networks obtained from Lennard-Jones clusters and clusters imposed on HCP lattices (to test influence of excluded volume on the results) as well as polybutadiene melts (to test the combined effect of excluded volume and chain connectivity). The network data are obtained as described below.

*Lennard-Jones Clusters* (*LJC*). The structure of clusters of atoms is an area of intense scientific research, since the properties of materials become size dependent when systems are small enough. By clusters, we refer to groups of atoms from tens to thousands of atoms. LJC are a group of atoms that contain purely Lennard-Jones interactions between pairs of atoms. Geometric optimization of these clusters requires developing efficient search algorithms, since the conformational space available to a cluster of atoms increases explosively. The atomic coordinates of LJC for sizes 3-1000 are deposited on the Cambridge Cluster Database [48]. Many of them are described by icosahedral motifs with an incomplete core [49]. Here we examine clusters of sizes 350 – 550, in intervals of 50 atoms. The cutoff distance for adjacency matrix construction is 1.6 Å [50]; see figure 1 for the RDF.

*Hexagonal Close Packed (HCP) lattice based atomic clusters:* We pack a set of *N*-atoms (nodes) on the lattice sites so that we have a finite system that has all lattice sites filled, unlike LJC that have incomplete cores. Although we studied the properties of simple cubic, body-centered cubic, face-centered cubic and HCP arrangements, here we present representative data from the latter only, as they all have similar results. In the HCP structure, nodes are arranged on a plane in a hexagonal formation, and planes are stacked on top of each other with alternating order. Although we display the RDF of this system in figure 1, we do not choose a cutoff distance where the first coordination shell ends, but we rather connect the first nearest neighbors to obtain the network; the fixed cutoff value is marked on the figure with the vertical dashed line. The generating function (equation 6) for $N = 500$ sites is $G_0(x) = 0.004x^3 + 0.032x^4 + 0.076x^5 + 0.028x^6 + 0.060x^7 + 0.080x^8 + 0.224x^9 + 0.048x^{10} + 0.064x^{11} + 0.384x^{12}$.

*Polybutadiene Melts* (*PBD*). We investigate networks constructed from PBD melts that have been obtained from molecular dynamics (MD) simulations. The system consists of monodisperse *cis*-1,4-PB of 32-chains, each with 32 repeat units ($C_{128}$). The initial coordinates of the system studied was prepared in Amorphous Construction Module of the Accelerys Material Studio 4.4 [51] at a density of 0.92 gr/cm$^3$, which occupies a cubic box of 47 Å on each side. Minimization, pre-equilibration and integration of the equations of motions were done with the NAMD program [52]. The interaction potentials for PBD chains reported in [53] are adopted. For all simulations, 1 fs integration time step was used. Temperature and pressure were maintained constant in the MD simulations at their prescribed values by employing the Langevin thermostat-barostat. For the non-bonding interaction cut-off distance of 10 Å was used with a switching function turned-on at 8 Å.

To obtain well-equilibrated samples of PBD chains with correct chain statistics, the initial structure which is energy minimized for 10000 steps is depressurized by placing the chains into a larger cubic box of 300 Å on each side. NVT simulations of this low-density system is carried out for 10 ns at 430 K. We then cool the system to 300 K by equilibrating for an additional 20 ns. Consequently, we compress it with NPT simulations at 1 atm at 430 K for 1 ns. We check that the conformational properties (as measured by the characteristic ratio) and the thermodynamic measurable (e.g. thermal expansion coefficient and compressibility) are compatible with the values in reference [53]. The data used in the current calculations are obtained from highly pressurized PBD melts via NPT simulations at 100 GPa and 430 K. We collect data for 50 ns. PBD melts are coarse grained by using the coordinates for the center of mass of carbon atoms in the butadiene repeat units. RDFs are obtained as usual, and cut-off distance for network construction is chosen at 5 Å, the ending point of the first coordination shell (figure 1).

## V. Statistical properties of close-packed atomic/molecular systems

The nearest neighbor degree correlations are displayed in figure 2 for the five systems studied. We find that all of them display assortative mixing. Furthermore, they are well-approximated by a linear relationship. In fact, one may use equation 14, which was obtained assuming that clustering is independent of degree, to predict the clustering coefficient (from the slope) and the ratio $<k^2>/z$ (from the intercept), to assess the range of validity of this assumption. In Table 1 is a comparative list of the predictions and the actual values calculated for the systems at hand. We find that the predictions overlap with the actual network values for all systems. Since the linear dependence, as well as the match between the predicted values of $C$ and $<k^2>/z$ depend on $C$ being independent of $k$ (see the reduction of eq. 13 to obtain eq. 14), we further examine this property in conjunction with degree distributions (figure 3). For all the systems studied, there is a decreasing trend of $C$ with $k$, although it is quite weak for RN, MN and LJC. Taken together with the degree distributions, also displayed in Figure 3, the variation of $C$ with $k$ is even less significant in the regions within one-standard deviation of the average degree for these three systems. Below we discuss in detail the implication of these observations for the individual systems studied.

**Self-organized molecular structures: Residue networks and micellar networks.** Previous studies on RN showed that these networks have high clustering as opposed to their random counterparts and have comparable shortest path lengths as the random networks; therefore, they can be considered as having small-world topology [16, 20-22]. In these studies, comparisons were performed for the average properties throughout the network between the RNs and their randomly rewired counterparts. Although average values do confirm that RNs have small-world properties, detailed analyses of the individual parameters are needed to assess similarity with artificially generated networks.

In reference [16] it was shown that the degree distributions of RN are Poisson; the mean is 6.2. Therein, it was also shown that the residues in the core have a mean clustering coefficient of ca. 1/3, whereas this value approaches 0.5 for the nodes that reside along the surface. Averaged over the set of 595 proteins, the clustering coefficient of RN has the value 0.38. The linearity between $k_{nn}$ and $k$ holds for all sizes of proteins, despite the size differences, in addition to the slight decreasing dependence of $C$ with $k$. We adopt equation 14 to analyze the relationship between $k_{nn}$ and $k$ in RN and we find that the slope can be identified by the average clustering coefficient of the network. The values of $<C>$ and $z=<k^2>/z$ calculated directly from the network and predicted via equation 14 are listed in Table 1. Within the error bounds, the predictions of theory are valid; the only slight deviation occurs as an underestimation of $<C>$ for the smaller proteins where the surface effects (and the variance in $C$) are more pronounced. We shall later elaborate further on the surface effects.

We expect other self-organized molecular structures to display network properties similar to the RN obtained from proteins, provided that they are thermodynamically stable and have a given average structure around which fluctuations are observed. Similar to the proteins, these structures follow certain organization rules due to the (in)compatibility of their chemical units with the solvent. Other environmental factors, such as the temperature or the concentration, play a role on the type of organization observed. As example systems, we choose micelles of different morphologies formed by the ABC type co-oligomers, whose coordinates are obtained from DPD simulations, as described in section IV.

At low concentrations, these oligomers organize to form spherical micelles. As the concentration increases, adjacent spheres merge and attain a cylindrical morphology. Further increase in the concentration results in the formation of lamellae. As an inset to figure 1, we display the spherical, cylindrical and the lamellar formations excerpted from oligomer concentrations of $v = 0.3$, 0.6, and 0.9, respectively. Note that it is the core region (i.e. the fluorinated regions shown as white spheres) that maintains the stable morphology, while the corona formed by the red and gray beads shows large fluctuations in conformation. Thus, we use the coordinates of the white blobs to generate the MN. The degree distribution and the dependence of clustering coefficient on degree of a sample network with $v = 0.6$ are shown in figure 3. It is important to note that, regardless of the type of self organization, these network parameters show the same pattern. We approximate their degree by Poisson distribution.

Similar to RN, analysis of $k$ vs. $k_{nn}$ relationship for MN reveals a positive linear correlation regardless of morphology (Figure 2). The values of $<C>$ and $z=<k^2>/z$ calculated directly from the network and predicted via equation 15 are also listed in Table 1. Nodes with less than four and more than 15 connections are omitted due to lack of statistics of blobs with so few or so many neighbors. Theoretical predictions of $z=<k^2>/z$ from the intercept of the $k$ vs. $k_{nn}$ relation is in excellent agreement with the numerical results. The slope of the best-fitting line slightly overestimates the average clustering coefficient.

The linear relationship between $k_{nn}$ and $k$ also predicts the increase in $z$ with size in RN and the decrease in $z$ with concentration (and morphology change) in MN. The theory slightly underestimates the clustering coefficient of RN whereas it overestimates that of MN. This is due to surface effects: In proteins, nodes along the surface have high clustering coefficients as shown in reference [16]. Because these nodes have few links that are interconnected, they increase the average clustering coefficient then would be directly predicted by an overall fit to the data in figure 2. Conversely, in MN surface nodes along the core are connected to the solvo-phillic arms of the chains. These connections, which are omitted in the calculations, since our network construction is based on only the core of the micelles and not the corona, have the reverse effect on the average value of the clustering coefficient.

**Effect of excluded volume: Lennard-Jones and HCP clusters.** Atoms or groups of atoms occupy a specific volume in space, and as a result, there is an upper bound on the number of neighbors that may be within the direct interaction range of a given node. Since our nodes comprise of coarse-grained groups of atoms that are not arranged spherically symmetric, we observe number of neighbors as large as 19 for a few nodes. This is in contrast to the maximum coordination of 12 expected of regular lattices of spherical particles. All of the networks studied here have this property of an upper bound on the degree. However, the extent to which this excluded volume effect influences the predictions of the previous subsections is unclear. To further investigate this point, we study LJC, which are clusters of atoms of minimum energy that interact purely via Lennard-Jones interactions. We confine out attention to those within the size range up to 550 which is compatible with the network sizes of RN and MN studied here. Although LJC conform to an icosahedral arrangement of atoms, they have incomplete cores (i.e. holes within the structure). We therefore also study hypothetical atomic clusters which have complete occupancy of HCP lattice sites.

The degree distributions of these systems are jagged and cannot be described as Poisson (figure 3). We find a linear relationship between $k_{nn}$ and $k$, as in the previous self-organized systems (figure 2). For LJC, the dependence of $C$ on $k$ is very similar to those of MN, following a nearly linear trend with a small negative slope (-0.02). For HCP, there is a stronger dependence of $C$ on $k$, yet for degrees that are observed more frequently, the average clustering remains almost constant ($C$ is 0.36 for $k = 12$ and 0.40 for $k = 9$). In both types of systems, while the $<k^2>/z$ values are well-predicted by equation 14, we find $<C>$ to be consistently underestimated by the theory, more so for LJC than for HCP (Table 1). As discussed in the previous subsection for RN, this is again due to the surface effects, which is more prominent for the irregular surfaces of LJC.

**Effect of chain connectivity: Polybutadiene Melts.** Finally, we study polymeric melts to discern the additional effect of connectivity on the statistical properties of the networks. The linear relationship between $k_{nn}$ and $k$ is also observed for this system which is forced into a close-packed structure by applying very high pressure. Degree distribution deviates from Poisson as for LJC and HCP, while clustering behavior is similar to those obtained for HCP. Both $<C>$ and $<k^2>/z$ are predicted via the theoretical fit (Table 1), with a slight overestimation of $<C>$. The overestimation is due to the fact that we truncate the system at the periodic boundaries of the cubic simulation box, and therefore the neighbors of some of the surface beads are artificially eliminated. Similar overestimation was also obtained for MN, where the corona neighbors of the core beads were removed. Thus, the effect of chain connectivity only plays a role in defining a correct neighborhood structure for the surface beads.

Putting together the results obtained thus far, we conclude that the excluded volume leads to the assortative mixing of the local structure, described by the positive slope of between $k_{nn}$ and $k$ curves. Furthermore, the extrapolation of the curves to low connectivity ($k \to 0$) leads to an excellent prediction of the $<k^2>/z$ values, regardless of the type of system studied (figure 4). Additional constraints on the local organization of the beads would lead to further local structuring which is measurable by the slope of these curves converging to $<C>$. We find that chain connectivity alone does not bring about such local organization of the beads as observed for PBD system at moderate density (data not shown). However, systems attaining dense core structures do converge to this limit. Such close-packing may be attained by imposing external factors such as the high pressure on PBD; alternatively, the core regions of self-organized systems prefer to realize such an arrangement due to the free energetic requirements of arranging chains with both solvo-phobic and solvo-phillic regions in a solvent that creates the driving force for the formation of the densely packed core [54].

## VI. Discussion

This study is based on the premise that network structures are better classified by the distributions of their network parameters rather than the average values. One previous example has been with approximating residue networks derived from proteins with the regular ring lattice: Although it is relatively easy to generate a corresponding ring lattice with few random rewired links having the same average degree and clustering coefficient as the RN [16], neither the second degree correlations nor the global properties (e.g. path length) are reproduced with this approach. However, comparison of distributions of the parameters involved is not straightforward.

To make the problem tractable, we derive a relationship between $k_{nn}$ and $k$ for networks with arbitrary degree distributions, but with narrowly distributed finite clustering. This subset of constraints is relevant to the study of complex systems, because the results directly apply to the study of self-organized molecular structures which are characterized by Poisson degree distributions, and narrowly distributed clustering coefficients. In randomly-packed chain systems this relationship is expected to be lost, as is observed when the corona region of the micellar networks (i.e. the disorganized parts of the chains protruding into the solvent) is also included in the calculations (data not shown). We validate the derived linear relationship between $k_{nn}$ and $k$ on several model networks based on three dimensional regular structures, polymeric melts forced into close-packing by external pressure as well as those constructed from proteins and micelles of self-organizing co-oligomers.

Excluded volume and close-packing together control the plateau value of the clustering coefficient reached for nodes which are located in the core of the systems studied; i.e. those with high degree. Moreover, they impose a decreasing trend on $C$ with increasing $k$, as well as providing restrictions on degree distributions. These constraints lead to assortative mixing in the graph structure. The presence of a single chain (as in RN), many chains (as in MN and PBD) or no chains (as in LJC and HCP) does not have an effect on these trends.

The close packed structures emerge as model systems that approximate the network properties of self-organized molecular structures: They yield the local statistical averages and distributions similar to that of the self-assembled systems. Using these model networks as the basis, one may generate novel networks by introducing a few random links whereby the local properties are preserved while the desired global properties are approximated. The ultimate goal is to use both statistical and spectral characterization to design networks with desired properties and to determine the principles underlying organization of complex networks.

# References

[1] B. A. Huberman, and L. A. Adamic, Nature **401**, 131 (1999).
[2] A. Vazquez, R. Pastor-Satorras, and A. Vespignani, Physical Review E **65**, 066130 (2002).
[3] D. J. Watts, and S. H. Strogatz, Nature **393**, 440 (1998).
[4] A. L. Barabasi *et al.*, Physica A **311**, 590 (2002).
[5] P. Erdös, and A. Rényi, Bulletin of the Institute of International Statistics (1961).
[6] A. L. Barabasi, and R. Albert, Science **286**, 509 (1999).
[7] J. Kleinberg, in *Proceedings of the thirty-second annual ACM symposium on Theory of computing* (ACM, Portland, Oregon, United States, 2000), pp. 163.
[8] G. Paperin, D. Green, and T. Leishman, 2008), pp. 575.
[9] R. Albert, and A. L. Barabasi, Reviews of Modern Physics **74**, 47 (2002).
[10] M. E. J. Newman, Siam Review **45**, 167 (2003).
[11] B. Bollábas, *Modern graph theory* (Springer, 1998).
[12] C. Godsil, and G. Royle, *Algebraic graph theory* (Springer, 2001).
[13] B. Mohar, Graph Theory, Combinatorics, and Applications **2** (1991).
[14] F. Chung, *Spectral graph theory* (AMS, 1997).
[15] M. Ipsen, and A. S. Mikhailov, Physical Review E **66**, 046109 (2002).
[16] A. R. Atilgan, P. Akan, and C. Baysal, Biophysical Journal **86**, 85 (2004).
[17] A. R. Atilgan, D. Turgut, and C. Atilgan, Biophysical Journal **92**, 3052 (2007).
[18] K. V. Brinda, and S. Vishveshwara, Biophysical Journal **89**, 4159 (2005).
[19] R. Sathyapriya, and S. Vishveshwara, Proteins **68**, 541 (2007).
[20] G. Bagler, and S. Sinha, Physica a-Statistical Mechanics and Its Applications **346**, 27 (2005).
[21] L. H. Greene, and V. A. Higman, Journal of Molecular Biology **334**, 781 (2003).
[22] M. Vendruscolo *et al.*, Physical Review E **65**, 061910 (2002).
[23] I. Bahar *et al.*, Physical Review Letters **80**, 2733 (1998).
[24] I. Bahar, A. R. Atilgan, and B. Erman, Folding & Design **2**, 173 (1997).
[25] I. Bahar *et al.*, Journal of Molecular Biology **285**, 1023 (1999).
[26] M. C. Demirel *et al.*, Protein Science **7**, 2522 (1998).
[27] V. A. Higman, and L. H. Greene, Physica a-Statistical Mechanics and Its Applications **368**, 595 (2006).
[28] C. Y. Lee, J. C. Lee, and R. R. Gutell, Physica a-Statistical Mechanics and Its Applications **386**, 564 (2007).
[29] U. K. Muppirala, and Z. J. Li, Protein Eng Des Sel **19**, 265 (2006).
[30] M. Aftabuddin, and S. Kundu, Biophysical Journal **93**, 225 (2007).
[31] C. Bode *et al.*, Febs Lett **581**, 2776 (2007).
[32] G. Amitai *et al.*, Journal of Molecular Biology **344**, 1135 (2004).
[33] C. Baysal, and A. R. Atilgan, Proteins-Structure Function and Genetics **45**, 62 (2001).
[34] A. del Sol *et al.*, Protein Science **15**, 2120 (2006).
[35] A. del Sol *et al.*, Mol Syst Biol, 19 (2006).
[36] K. V. Brinda, and S. Vishveshwara, Bmc Bioinformatics **6**, 296 (2005).
[37] R. Pastor-Satorras, A. Vazquez, and A. Vespignani, Physical Review Letters **87**, 258701 (2001).
[38] M. E. J. Newman, Physical Review E **67**, 026126 (2003).
[39] R. Xulvi-Brunet, and I. M. Sokolov, Physical Review E **70**, 066102 (2004).
[40] M. E. J. Newman, Physical Review Letters **89**, 208701 (2002).
[41] M. E. J. Newman, S. H. Strogatz, and D. J. Watts, Physical Review E **6402**, 026118 (2001).

[42]  Á. Serrano, and M. Boguñá, Physical Review E **72**, 036133 (2005).
[43]  H. M. Berman *et al.*, Acta Crystallogr D **58**, 899 (2002).
[44]  P. Fariselli, and R. Casadio, Protein Engineering **12**, 15 (1999).
[45]  C. Atilgan, O. B. Okan, and A. R. Atilgan, Proteins: Structure, Function, Bioinformatics, accepted.
[46]  A. S. Ozen, U. Sen, and C. Atilgan, J Chem Phys **124**, 064905 (2006).
[47]  G. Kacar, C. Atilgan, and A. S. Ozen, J Phys Chem C **114**, 370 (2010).
[48]  J. P. K. D. D. J. Wales, A. Dullweber, M. P. Hodges, F. Y. Naumkin F. Calvo, J. Hernández-Rojas and T. F. Middleton.
[49]  Y. H. Xiang *et al.*, J Phys Chem A **108**, 3586 (2004).
[50]  Y. H. Xiang *et al.*, J Phys Chem A **108**, 9516 (2004).
[51]  Accelrys Inc.,  (Accelrys Inc., San Diego, 2008).
[52]  J. C. Phillips *et al.*, J Comput Chem **26**, 1781 (2005).
[53]  G. Tsolou, V. A. Harmandaris, and V. G. Mavrantzas, Macromol Theor Simul **15**, 381 (2006).
[54]  H. Can, G. Kacar, and C. Atilgan, J Chem Phys **131**, 124701 (2009).

**Table 1.** Network parameters $<C>$ and $<k^2>/z$ computed from the generated graphs and predicted from the least squares linear fit to $k_{nn}$ vs. $k$ curves.

|  |  | Calculated | | Predicted[c] | |
|---|---|---|---|---|---|
|  |  | $<C>$ | $<k^2>/z$ | $<C>$ | $<k^2>/z$ |
| **Residue Networks**[a] | 595 Proteins; $<N>$ = 254 | 0.38 | 6.2 | 0.35±0.01 | 5.8±0.2 |
|  | $N$ = 140-160 | 0.38 | 6.1 | 0.32±0.01 | 5.7±0.2 |
|  | $N$ = 190-210 | 0.39 | 6.2 | 0.32±0.02 | 5.8±0.4 |
|  | $N$ = 290-310 | 0.37 | 6.6 | 0.36±0.01 | 6.2±0.2 |
| **Micellar Networks**[a] | $v$ = 0.3 | 0.45 | 10.3 | 0.40±0.02 | 10.5±0.8 |
|  | $v$ = 0.6 | 0.43 | 9.9 | 0.51±0.02 | 10.2±0.8 |
|  | $v$ = 0.9 | 0.41 | 9.4 | 0.51±0.02 | 9.6±0.6 |
| **Lennard-Jones Clusters**[b] | $N$ = 350 | 0.47 | 15.1 | 0.33±0.07 | 14.4±1.4 |
|  | $N$ = 400 | 0.47 | 15.3 | 0.31±0.06 | 14.5±1.1 |
|  | $N$ = 450 | 0.46 | 15.4 | 0.33±0.07 | 14.6±1.3 |
|  | $N$ = 500 | 0.46 | 15.5 | 0.33±0.07 | 14.6±1.4 |
|  | $N$ = 550 | 0.47 | 15.6 | 0.37±0.12 | 15.3±2.6 |
| **HCP**[b] | $N$ = 500 | 0.41 | 10.2 | 0.38±0.06 | 9.9±0.8 |
| **PBD**[b] | $T$ = 430 K, $P$ = 100 GPa | 0.45 | 12.8 | 0.52±0.03 | 12.4±0.7 |

[a] Degree distribution is well-described by Poisson; therefore predictions by eq. 14 and 15 lead to the same result. Also $z = <k> = <k^2>/z$ for these systems.
[b] Degree distributions are not well-described by Poisson. Predictions are made through eq. 14.
[c] Error margins on the predicted values are reported.

**Figure Captions:**

**Figure 1**. Radial distribution function $g(r)$ calculated for sample systems in the current work. Distance $r$ is in Å for RN, PBD and LJC structures, and is in reduced units (bead size = 1 unit) for the other cases. The cutoff distances, $r_c$, utilized for network construction are also marked on the figures. An example network construction is displayed for the residue network (RN) of the sample protein (PDB code **1esl**) as an inset; protein structure in ribbon diagram is on the left, the constructed network at the $r_c$ value selected for all residue networks is on the right. Also shown as inset are the MN structures formed at various concentrations ($\nu = 0.30$, spherical; $\nu = 0.60$ cylindrical; $\nu = 0.90$, lamellar).

**Figure 2.** Averaged $k_{nn}$ vs. $k$ plots for RN with $N = 190\text{-}210$ (29 proteins), MN with $\nu = 0.60$ (cylindrical micelle is formed in the core), LJC ($N = 500$), HCP ($N = 500$), and PBD systems. Using equation 14, the values for $C$ and $z$ are predicted and compared with the actual values of the network in Table 1. For RN, nodes with degree 1, 13, 14 and 15 are omitted since there is relatively small number of nodes with such degrees (< 25) to provide meaningful statistics. For MN, nodes with degree less than 5 and greater than 15 are omitted to provide meaningful statistics.

**Figure 3.** Averaged clustering vs. degree plots for RN ($N = 190\text{-}210$), MN ($\nu = 0.60$), LJC ($N = 500$), HCP ($N = 500$), and PBD on the left y-axis. Degree distributions are superposed (shaded) and labeled on the right y-axis.

**Figure 4.** Comparison of predicted versus calculated values of the ratio of second to first moments of the degree distributions, $<k^2>/z$.

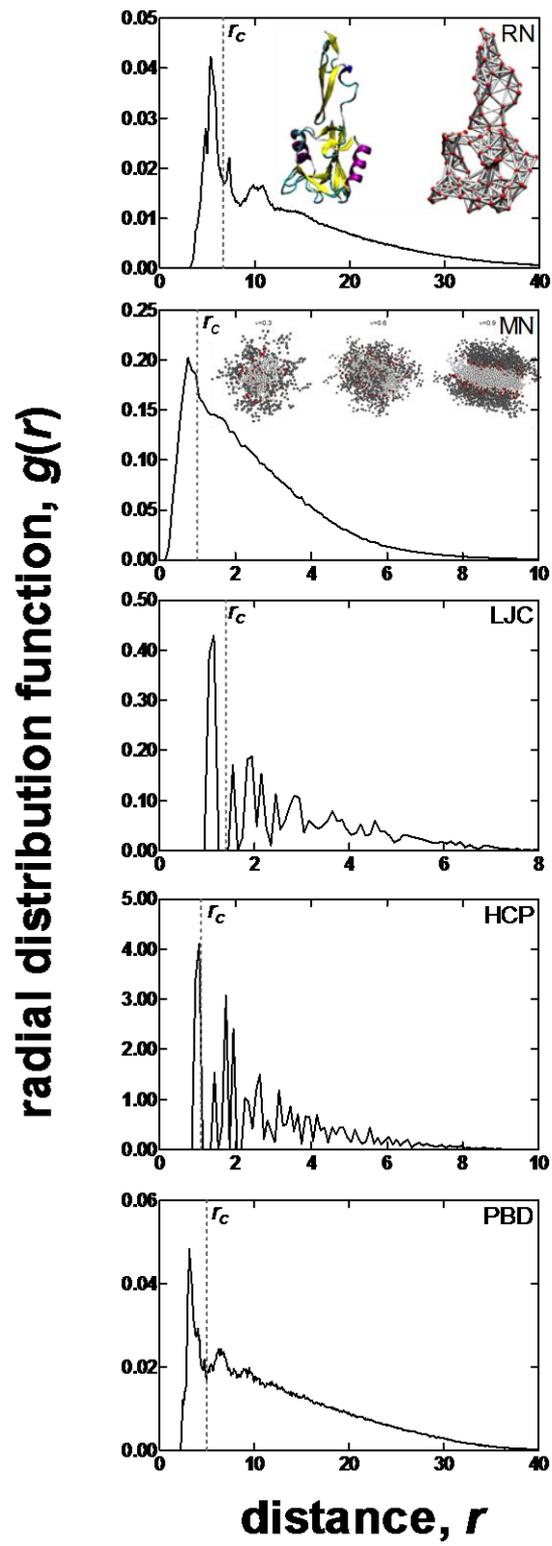

*Figure 1*

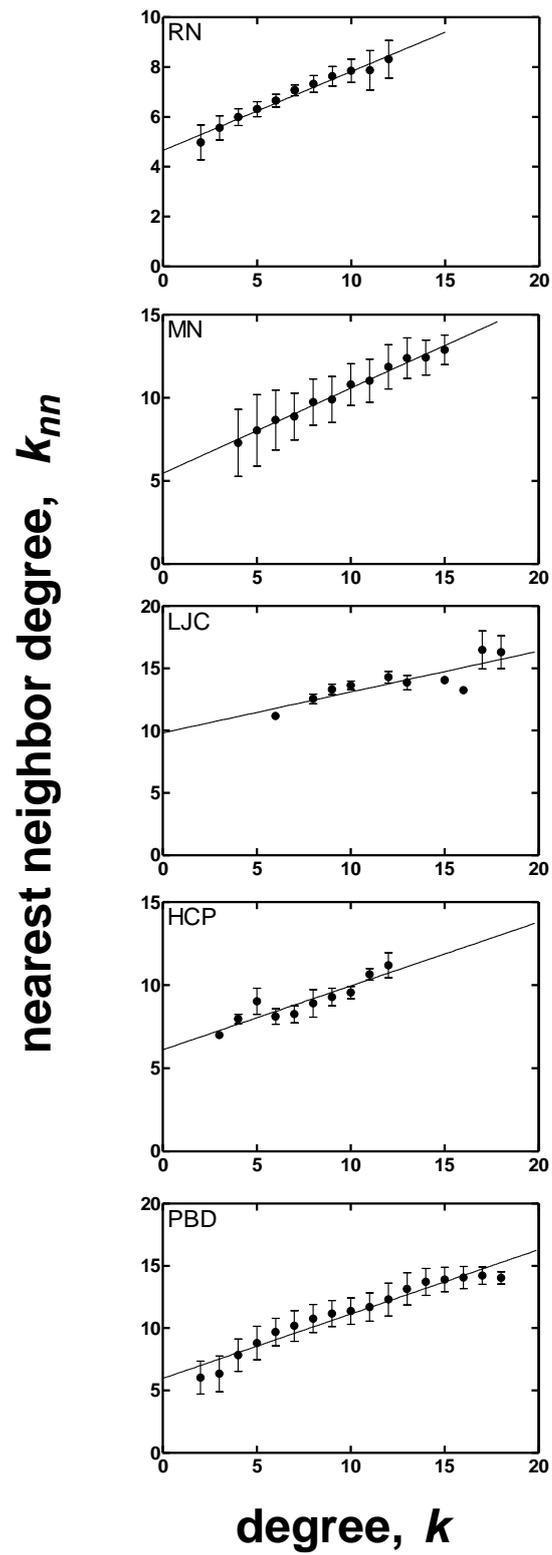

**Figure 2**

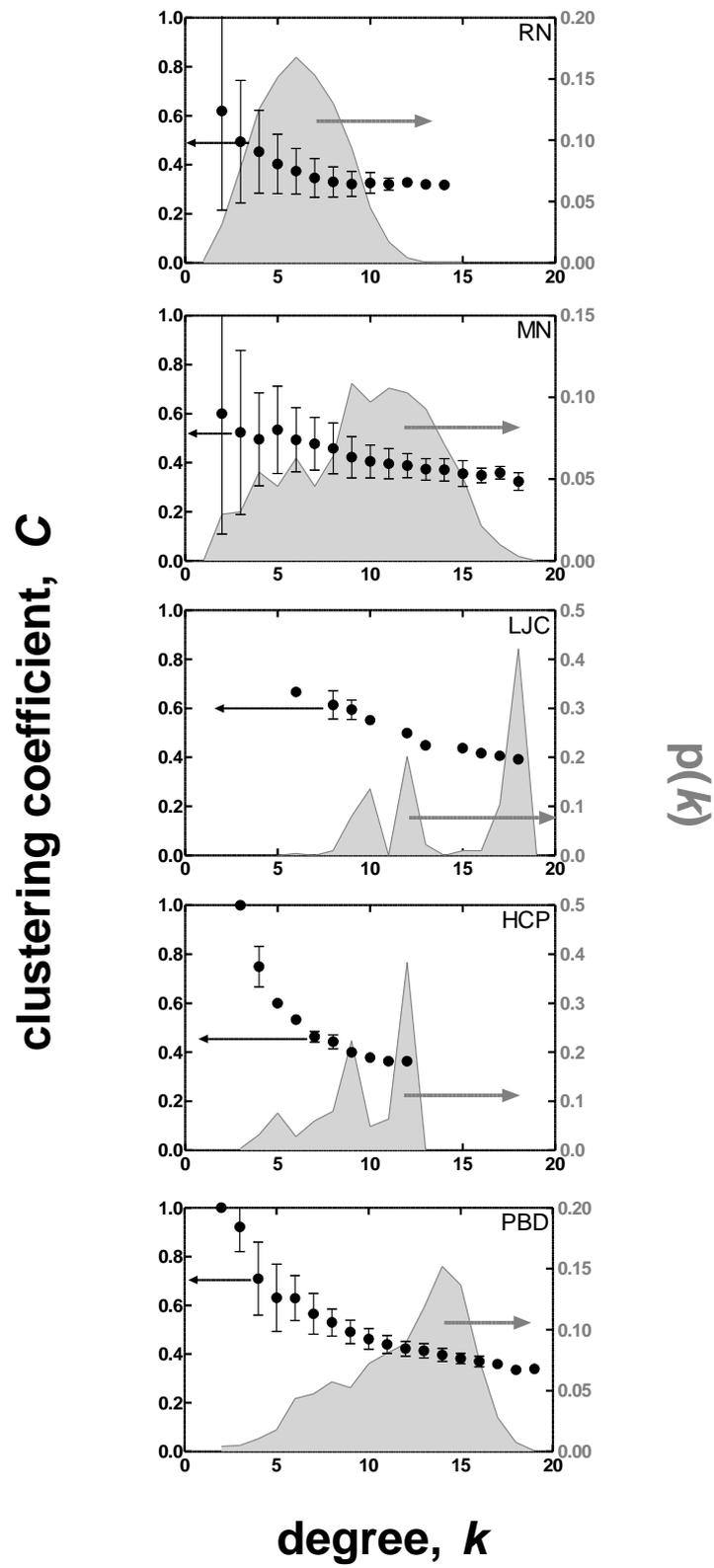

**Figure 3**

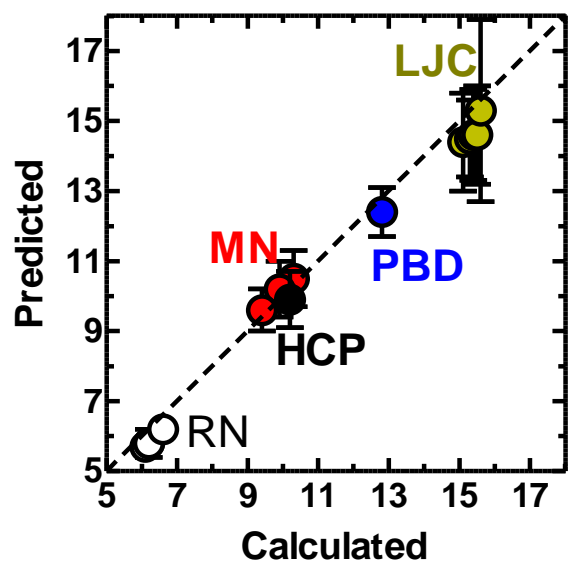

**Figure 4**